\documentclass[
amsmath,amssymb,
aps,
prl,
twocolumn, superscriptaddress, showpacs
]{revtex4-1}

\usepackage{graphicx}
\usepackage{dcolumn}
\usepackage{bm}
\usepackage{mathtools}
\usepackage{xcolor}

\graphicspath{{pics/}}

\def \bra#1{\mathinner{\langle{#1|}}}
\def \ket#1{\mathinner{|{#1}\rangle}}

\def \red #1 {\textcolor{red}{#1}}

\usepackage{tikz}
\usepackage{circuitikz}
\usetikzlibrary{backgrounds,fit,decorations.pathreplacing,calc,backgrounds}  

\begin{document}


\title{Quantum Gradient Algorithm for General Polynomials}

\author{Keren Li}
\thanks{These authors contributed equally to this work.}
\affiliation{State Key Laboratory of Low-Dimensional Quantum Physics and Department of Physics, Tsinghua University, Beijing 100084, China}
\affiliation{Center for Quantum Computing, Peng Cheng Laboratory, Shenzhen 518055, China}

\author{Pan Gao}
\thanks{These authors contributed equally to this work.}
\affiliation{State Key Laboratory of Low-Dimensional Quantum Physics and Department of Physics, Tsinghua University, Beijing 100084, China}

\author{Shijie Wei}
\affiliation{Beijing Academy of Quantum Information Sciences, Beijing 100193, China}
\affiliation{State Key Laboratory of Low-Dimensional Quantum Physics and Department of Physics, Tsinghua University, Beijing 100084, China}

\author{Jiancun Gao}
\affiliation{State Key Laboratory of Low-Dimensional Quantum Physics and Department of Physics, Tsinghua University, Beijing 100084, China}

\author{Guilu Long}
\email{gllong@tsinghua.edu.cn}
\affiliation{State Key Laboratory of Low-Dimensional Quantum Physics and Department of Physics, Tsinghua University, Beijing 100084, China}
\affiliation{ Beijing National Research Center for Information Science and Technology and School of Information Tsinghua University, Beijing 100084, China}
\affiliation{Beijing Academy of Quantum Information Sciences,  Beijing 100193, China}
\affiliation{Frontier Science Center for Quantum Information, Beijing 100084, China}

\date{\today}

\begin{abstract}
Gradient-based algorithms, popular strategies to optimization problems, are essential for many modern machine-learning techniques. Theoretically, extreme points of certain cost functions can be found iteratively along the directions of the gradient. The time required to calculating the  gradient of $d$-dimensional problems is at a level of $\mathcal{O}(poly(d))$, which could be boosted by quantum techniques, benefiting the high-dimensional data processing, especially the modern  machine-learning engineering with the number of optimized parameters being in billions.
Here, we propose a quantum gradient algorithm for optimizing general polynomials with the dressed amplitude encoding, aiming at solving fast-convergence polynomials problems within both time and memory consumption in $\mathcal{O}(poly (\log{d}))$. Furthermore, numerical simulations are carried out to inspect the performance of this protocol by considering the noises or perturbations from initialization, operation and truncation. For the potential values in high-dimension optimizations, this quantum gradient algorithm is supposed to facilitate the polynomial-optimizations, being a subroutine for future practical quantum computer.
\end{abstract}
\maketitle


\section{\label{sec1:level1}Introduction}
Recent advances indicate that machine learning(ML) methods is coming into prominence as the potential solutions to various challenging physical problems, such as identification of phases of matter\cite{2016-wang-MLphases,2017-Van-Learningphases}, representation of many-body states\cite{2017-Troyer-NNquantum,2011-Ulrich-TNquantum} and quantum entanglement, tomography\cite{2017-Deng-NNentanglement,2019-Levine-learningEnt,2018-Troyer-NNtomography}. Gradient-based algorithms, consisting of the prototypical gradient method and its variants, are essential to many optimization problems, which are  the keys to most ML methods. They are extensively applied to logistic regression, support vector machine, neural network and those whose inside parameters are to be optimized\cite{2019-Zhang-GforDP,2018-Manogaran-GLR,2018-wang-GSVD,2018-Du-GNN}. With the advent of vigorous machine-learning(ML) methods, which have been ubiquitously powering the modern technologies\cite{2014-Sun-facerecog,2011-Ricci-recom,2016-Boj-SelfD}, ML methods are destined to process an incredible amount of data generated in this internet era. Although, the time required to calculating the  gradient of $d$-dimensional problems is at a level of $\mathcal{O}(poly(d))$. For the modern architecture, the number of parameters to be optimized would be in billions. Calculating such a scale of gradient would be computationally intractable.

Though to-date quantum computing engineering is still on a modest stage which is far to meet the requirements of practical quantum computing, dramatic hardware improvements has been achieved for decades in both the scale and the quality of qubits\cite{2019-google-Qsupremacy,2020-honeywell-ion}. It offers a tantalizing prospect to outperform the most powerful electronic computer\cite{2002-nielson-QCQI}, providing an exponential speed-up for certain problems\cite{2017-Seth-Nat-quantum,2020-google-TFQ}. Many quantum enhanced algorithms are emerging, such as quantum Fourier transformation(QFT)\cite{2002-nielson-QCQI}, $HHL$ linear equation algorithm\cite{2009-harrow-PRL-HHL} and quantum principle component analysis(qPCA)\cite{2014-Seth-Np-quantum}. As for enhancing the gradient-based algorithms, plenty of research on their quantum version has been published\cite{2005-jordan-Qgradient,2019-Nathan-qGradient,2017-Hybrid-Li,2019-qGradient-schuld,2019-Rebentrost-Qgradient,2020-qGradient-Kerenidis}. However, as their different applicable conditions, a general quantum gradient algorithm is still required.

As polynomials can not only be directly applied into some ML models but also be approximation of arbitrary functions\cite{2003-oh-PNN1,2015-Zjavka-PNN2}, in this paper, we propose a quantum gradient algorithm correspondingly based on the previous framework\cite{2019-Rebentrost-Qgradient}. The dressed amplitude encoding(DAE) method and a non-unitary subroutine are introduced, realizing the calculation on the gradient of general polynomials which is no more homogeneous and even order. As constrained optimization problems usually can be transformed into unconstrained optimization problems by introducing penalty function or Lagrange method, our protocol, dropping the constraints on feasible points, extends the framework to more optimization cases. Besides, numerical simulations are conducted within consideration of noise or perturbations  for both maximum and minimum problems. Although it would be more applicable in fast convergence problems due to the finite success probability and the multi-copies required for each iteration. By the results, our adapted protocol shows the robustness to experiment concerned errors of reasonable strengths, which is important for the real application in Noisy Intermediate-Scale Quantum (NISQ) computers. Moreover, the protocol inherits the advantage of the amplitude encoding that reducing both memory and time consumption to $\mathcal{O}(poly\log{d})$. As the importance of high dimensional optimization problems in modern machine-learning methods, this algorithm have the potential to boost the interdisciplinary research of quantum computing and machine-learning.

\section{\label{result}Result}
Maximizing or minimizing $f(\bm{x})$, where $\bm{x}$ is a $d$-dimensional real variable $(x_1,x_2,\cdots,x_d)^T$, is a prototypical optimization problem and gradient-based algorithms are usually resorted to. Let $\xi$ be the  learning rate and superscript $t$ stands for the iteration steps, the variable thus can be updated iteratively with
\begin{eqnarray}
\bm{x}^{t+1}=\bm{x}^t \pm \xi \bm{\nabla} f(\bm{x}^t).
\end{eqnarray}
where $+(-)$ corresponds to the maximum(minimum) problem. To be noticed, all the superscript(and subscript) denoting step index and variable will be omitted in remaining text. 

\emph{\textbf{Protocol}---}
General polynomials optimization with no more than $2p$-order is considered in this paper. By introducing $\bm{X}=(1,\bm{x}^T)^T$, polynomial cost function can be written as
\begin{eqnarray} \label{classical_f}
 f(\bm{X})=\frac{1}{2} \bm{X}^{T \otimes p} A  \bm{X}^{\otimes p}
\end{eqnarray}
where $A$ is a $(d+1)^{p} \times (d+1)^{p}$ matrix which specifies the coefficient of polynomials. Inasmuch as complex number can be treated as two independent real ones, and global scaling of $f$ cause no impact on the extreme points, $A$ can always be chosen as real symmetric and scaled as $A\sim A/p||A||_{max}$ for convenience of following analysis.

The corresponding gradient $\bm{\nabla}f$ can be expressed as a part of $\hat{D}\bm{X}$ as
\begin{eqnarray}\label{classical_d}
\begin{pmatrix}
\kappa \\
\bm{\nabla}f
\end{pmatrix}
=\left[\bm{X}^{T\otimes p-1}\sum_{k=1}^pP_k A P_k^{\dagger}\bm{X}^{\otimes p-1}\right] \bm{X}\equiv\hat{D}\bm{X} ,
\end{eqnarray}
where $P_k$ is the permutation operator, swapping the $1$-st and the $k$-th $(d+1) \times (d+1) $ subspaces in $A$. Thus $\hat{D}=\bm{X}^{T\otimes p-1}\sum_{k=1}^pP_k A P_k^{\dagger}\bm{X}^{\otimes p-1}$ is dubbed as a variable dependent gradient operation. Noticeably, $\kappa$ is redundant and is to be cast away to export a right $\bm{\nabla}f$ with the expression of $\hat{D}$. 

Therefore, Eq.(\ref{classical_f}) and Eq.(\ref{classical_d}) establish the depiction of the polynomials optimization. As for its quantum version, a dressed amplitude encoding(DAE) method introduced as 
\begin{eqnarray}
\bm{X} \rightarrow \ket{\bm{X}}=\cos{\gamma} (\ket{0}+ \sum_{i=1}^d x_i \ket{i}),
\end{eqnarray}
with the $\bm{x}$-dependent normalization factor $\cos\gamma$ that satisfy $1/\cos^2\gamma$=$||\bm{X}||^2$. The DAE inherits the pleasing feature of resources saving as in the previous work\cite{2019-Rebentrost-Qgradient}.  Moreover, the quantum counterparts of Eq.(\ref{classical_f}) and Eq.(\ref{classical_d}) can be 
\begin{eqnarray}\label{quant_class}
\cos^{2p}\gamma f(\bm{X})&=&\frac{1}{2}\bra{\bm{X}}^{\otimes p} A \ket{\bm{X}}^{\otimes p}\nonumber \\
\cos^{2p-1}{\gamma}\hat{D}\bm{X}&=& Tr_{p-1}[\rho_{\bm{X}}^{\otimes p-1} \mathcal{M}] \ket{\bm{X}}=\mathcal{D} \ket{\bm{X}}
\end{eqnarray}
where $\rho_{\bm{X}}=\ket{\bm{X}}\bra{\bm{X}}$ and $\mathcal{M}=\sum_{k=1}^p P_k A P_k$. Therefore, the iterative process of the gradient-based method is implemented as
\begin{eqnarray}\label{quantum_update}
\ket{\bm{X}'}=\cos{\gamma'} (\ket{0} +\sum_{i=1}^d x'_i \ket{i}) \propto\ket{\bm{X}}\pm \xi\mathcal{K} \mathcal{D} \ket{\bm{X}},
\end{eqnarray}
This equation implies that $\cos{\gamma}^{2p-2}$ is absorbed into $\xi$, as the adjustable learning rate. Besides, the non-unitary operator $\mathcal{K}=diag(0,1,1,\cdots,1)$ is introduced after the application of $\mathcal{D}$, discarding the $\ket{0}$ component in $\mathcal{D}\ket{\bm{X}}$ and avoiding a trapped fake optimized state $\ket{\bm{X}^{trap}}$,  which satisfies $\mathcal{D}\ket{\bm{X}^{trap}}\propto\ket{\bm{X}^{trap}}$.
In this way, the final optimized $\ket{\bm{X}^{op}}$ is obtained and $\cos{\gamma^{op}}$ can be estimated within relative error $\epsilon_\gamma$ by consuming of $\mathcal{O}(\epsilon_{\gamma}^2)$ copies of this final states\cite{supp}. Hence, classical results can be got by reversing the DAE. 

Noticeably, the combination of DAE and $\mathcal{K}$ provides an alternative way to calculate the gradient of a general polynomials, varying from the previous work dealing with constrained homogeneous optimization. As some constrained optimization problems can be transformed into unconstrained optimization problems by introducing penalty function or Lagrange method, it extends the framework to more optimization cases. 
 
\begin{figure}[!h]
  \centerline{
  \resizebox{1\columnwidth}{!}{%
  \begin{tikzpicture}[thick]
    \tikzstyle{operator} = [draw,fill=white,minimum size=1.5em] 
    \tikzstyle{operator2} = [draw,fill=white,minimum size=4em] 
    \tikzstyle{operator3} = [draw,shape=rectangle,fill=yellow!35,minimum width=1cm, minimum height = 0.7cm] 
    \tikzstyle{phase} = [fill,shape=circle,minimum size=3pt,inner sep=0pt]
    \tikzstyle{phase0} = [draw, fill=white,shape=circle,minimum size=3pt,inner sep=1pt]
    \tikzstyle{ellipsis} = [fill,shape=circle,minimum size=2pt,inner sep=0pt]
    \tikzset{meter/.append style={fill=white, draw, inner sep=5, rectangle, font=\vphantom{A}, minimum width=15, 
 path picture={\draw[black] ([shift={(.05,.2)}]path picture bounding box.south west) to[bend left=40] ([shift={(-.05,.2)}]path picture bounding box.south east);\draw[black,-latex] ([shift={(0,.15)}]path picture bounding box.south) -- ([shift={(.15,-.08)}]path picture bounding box.north);}}}
    \node at (0,0) (q1) {$\ket{0}_k$};
    \node at (0,-1) (q2) {$\ket{0}_{up}$};
    \node at (0,-2) (q3) {$\ket{0}_{d}$};
    \node at (0,-3) (q4) {$\ket{0}_e$};
    \node at (0,-4) (q5) {$\ket{\bm{X}}_v$};
    \node[meter] (end1) at (11.5,0) {} edge [-] (q1);
    \node[meter]  (end2) at (11.5,-1) {} edge [-] (q2);
    \node[meter]  (end3) at (11.5,-2) {} edge [-] (q3);
    \node (end4) at (11.5,-3) {} edge [-] (q4);   
     \node[meter]  (end5) at (11.6,-4) {} edge [<-] (q5);
    \node (sl41) at (0.65,-3) {/} ;
    \node (sl41) at (0.65,-4) {/} ;
    \node[operator] (op2) at (1.5,-1) {$R_x(\eta)$} ;
    \node[operator] (op2) at (2.5,-3) {H} ;
    \node[phase] (phase11) at (3,-1) {};
    \node[phase] (phase12) at (3,-3) {};
    \node[operator3] (op21) at (3,-4) {$e^{-i\mathcal{D}t}$} ;
    \draw[-] (phase11) -- (phase12);
    \draw[-] (phase12) -- (op21);  
    \node[phase] (phase13) at (3.6,-1) {};
    \node[operator] (op23) at (3.6,-3) {$U^{-1}_{F}$} ;
    \draw[-] (phase13) -- (op23);
    \node[operator] (op1) at (5,-2) {$R_x(\theta)$};
    \node[phase] (phase21) at (5,-1) {};
    \node[phase] (phase22) at (5,-3) {};
    \draw[-] (op1) -- (phase21);
    \draw[-] (op1) -- (phase22);
    \node[phase] (phase61) at (6.4,-1) {};
    \node[operator] (op61) at (6.4,-3) {$U_{F}$} ;
    \draw[-] (phase61) -- (op61);
    \node[phase] (phase71) at (7,-1) {};
    \node[phase] (phase72) at (7,-3) {};
    \node[operator3] (op71) at (7,-4) {$e^{i\mathcal{D}t}$} ;
    \draw[-] (phase71) -- (phase72);
    \draw[-] (phase72) -- (op71);
        \node[operator] (op75) at (7.5,-3) {H} ;   
        \node[operator] (op75) at (8.5,0) {H} ;
     \node[phase] (phase81) at (9,-0) {} ;
     \node[phase] (phase82) at (9,-1) {} ;
     \node[phase0] (phase83) at (9,-2) {} ;
     \node[operator] (op81) at (9,-4) {$E$} ;
     \draw[-] (phase81) -- (phase83);
     \draw[-] (phase83) -- (op81);
        \node[operator] (op85) at (9.5,0) {H} ;
        \node[operator] (op85) at (10.5,-1) {$R_x(\eta)$} ;
     \node at ($(end5)+(0.1,-0.7)$){$\ket{\bm{X}'}_v$};
     \node at ($(0,-5)$){\emph{\textbf{Stage-1}}};
     \node at ($(5,-5)$){\emph{\textbf{Stage-2}}};
     \node at ($(9.5,-5)$){\emph{\textbf{Stage-3}}};
  \begin{pgfonlayer}{background} 
  \node[rectangle, fill=blue!10, rounded corners=1mm, minimum height=4.7cm, opacity=.8, text width=0.7cm, align=left, outer sep=5mm] (main) at (-0.1,-2) {};
  \node[rectangle, fill=pink!35, rounded corners=3mm, minimum height=3cm, opacity=.9, text width=5.5cm, align=left, outer sep=5mm] (main) at (5,-3) {};
   \node[rectangle, fill=green!20, rounded corners=3mm, minimum height=4.7cm, opacity=.9, text width=1.5cm, align=left, outer sep=5mm] (main) at (9,-2.0) {};
  \end{pgfonlayer} 
 \end{tikzpicture} }
 }
 \caption{Quantum circuit to implement the protocol, which consists of three stages (dashed lines) that evolving the system to the current iterative state. Four subroutines which implement initialization, non-unitary $\mathcal{D}$(by \emph{HHL}-like methods]), evolution $e^{-i\mathcal{D}t}$(by \emph{qPCA} and \emph{Hamiltonian simulation} methods) and $\mathcal{K}$ truncation  are labeled in blue, pink, yellow and green in sequence. }   
  \label{algorithm_circuit} 
\end{figure}
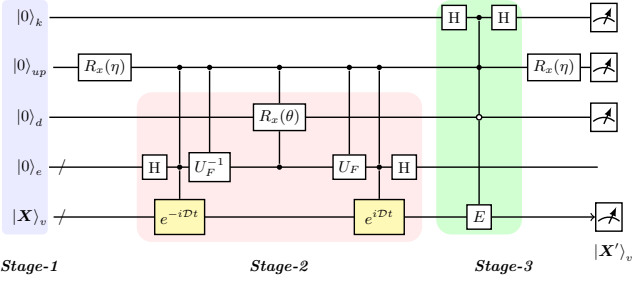

\emph{\textbf{Circuit}---} 
Circuit to implement Eq.(\ref{quantum_update}) is specified in (Fig.\ref{algorithm_circuit}). It consists of three stages (divided by dashed lines), including four subroutines(colored), where, the \emph{blue} block denotes the initialization of the input, the \emph{pink} one denotes the construction of variable dependent $\mathcal{D}$, the \emph{yellow} one denotes the construction of $e^{-i\mathcal{D}t}$ and the \emph{green} one denotes the construction of $\mathcal{K}$. In details, the last three block are realized with the assistance of the qPCA\cite{2014-Seth-Np-quantum}, \emph{$HHL$-like methods}\cite{2009-harrow-PRL-HHL,wei2017realization} and \emph{linear combinations of unitary operators}\cite{2011-Long-DQC,2020-Wei-FQE,2010-Achilds-HSimulation,2015-DBerry-HSimulation}.

For implementation, $5$ principle registers are required. Three are labeled as $k,up,d$ with one-qubit, assisting the construction of $\mathcal{K}$, variable updating and $\mathcal{D}$. One with $\chi$ qubits serves as the storage for eigenvalues of $\mathcal{D}$(labeled as $e$). The final register with $\log{{(d+1)}}$ qubits is labeled as $v$, encoding the variable $\ket{\bm{X}}$. Besides, there are implicit requirements of $[m(p-1)+1]$ copies of $\ket{\bm{X}}$ for implementation of variable dependent $e^{-i\mathcal{D}t}$. All $\chi, m$ determined by wanted precision and will be specified later.

In the following, we present a sketch of the protocol. Complexity analysis, as well as the error analysis is also given thereafter with the reasonable assumptions.

\emph{Stage-1:}
\emph{Subroutine-1}(colored blue) is the main body of this stage which encodes the current variable  $\bm{X}$ as $\ket{\bm{X}}_v$, i.e.
\begin{eqnarray}
\ket{0}_{k}\ket{0}_{up}\ket{0}_{d}\ket{0}_e \ket{\bm{X}}_v,
\end{eqnarray}

Typically, the concerned optimal problems are usually insensitive to the initial variables. As is verified in numerical simulation, we can start with some easy-access states. Thus, the complexity of this subroutine can be ignored in the first iteration step. As for the following iterations, output of the former iteration can be used as the input. Even for a general $\bm{X}$, it can be realized via quantum random access memory(qRAM)\cite{2008-giovannetti-Qram,2008-giovannetti-ArchitectureQram} or Hamiltonian simulation like method\cite{2018-Wossning-QLSA}. 

Therefore, a conservative estimate of \emph{Subroutine-1} is admitted according to the qRAM or Hamiltonian simulation within $\mathcal{O}(poly\log d)$ time complexity\cite{supp}.

\emph{Stage-2:}
a \emph{HHL}-like subroutine(\emph{Subroutine-2}) for implementing the non-unitary operation $\mathcal{D}$, and a \emph{qPCA} $\&$ \emph{Hamiltonian simulation} subroutine(\emph{Subroutine-3}) for implementing the variable-dependent control evolution $C_{\mathcal{U_D}}$=$\sum_{j=0}^{2^\chi-1} \ket{j}\bra{j} e^{-i \mathcal{D} \frac{2\pi}{2^\chi} j}$, are involved in this part.

Local operation ${R^x_{up}(\eta)}$ is a preliminary treatment, which rotates $|0\rangle_{up}$ to $\cos{\eta}\ket{0}_{up}+\sin{\eta}\ket{1}_{up}$. Then a \emph{HHL}-like method (colored pink) is implemented with the access to Hadamard gate $H_e$, controlled evolution  $C_{\mathcal{U_D}}$, controlled Fourier transform $C_{\mathcal{U}_F}$ and multi-controlled rotation $C_{R^x(\theta)}$, producing
\begin{eqnarray}
\!\cos{\eta}\!\ket{0}\!_{up}\!\ket{0}_{d}\!\ket{\bm{X}}_v\!+\!\sin{\eta}\!\ket{1}\!_{up}\!(\ket{0}\!_{d}\mathcal{D}\!\ket{\!\bm{X}}\!+\!\ket{1}\!_{d}\mathcal{D}\!^{\perp} \!\ket{\!\bm{X}}\!)\quad
\end{eqnarray}
where $\mathcal{D}^{\perp}\ket{\bm{X}}=\sum_k \sqrt{1-\lambda^2_k} \beta_k \ket{k}$ and the register $e$ is formatted via uncomputing.

On the premise that $C_{\mathcal{U_D}}$ is easy to realize, the complexity of this stage mainly comes from quantum phase estimation(qPE), which is specified in the standard textbook, requiring $\chi=n_p+ \ulcorner log(2+\frac{1}{2\delta})\urcorner$ qubits and $\mathcal{O}(\chi^2)$ elemental gates. This guarantees a $1-\delta$ success probability and bounded binary estimating error $\epsilon_p=2^{-n_p}$. However, implementing $C_{\mathcal{U_D}}$ is nontrivial for its variable-dependence. We turn to \emph{Subroutine-3} which integrates both qPCA method and Hamiltonian simulation.
In this way, $C_{\mathcal{U_D}}$ is realized with a bounded error $\epsilon_{\mathcal{D}}=\epsilon_{pca}+\epsilon_{hs}$. As to the circuit depth for $C_{\mathcal{U_D}}$, it requires $\mathcal{O}(2 p\pi s||A||_{max}+\frac{ \log{1/\epsilon_{hs}}}{\log{\log{1/\epsilon_{hs}}}})$ queries on the coefficient oracles of $A$ and $\mathcal{O}(4 \pi^2 p^3 ||A||^2_{max} \log{(1+d)}/{\epsilon_{pca}})$ 2-qubit swap gates($s$ is the sparsity of $A$). With respect to the circuit size,  it requires $\mathcal{O}({4 \pi^2 p^3 ||A||^2_{max}}/{\epsilon_{pca}})$ copies for state $\ket{\bm{X}}$. Details on subroutines can be found in the Appendix\cite{supp}. 

\emph{Stage-3:} 
For removing the $\ket{0}$ component in $\mathcal{D}\ket{\bm{X}}$ and avoiding 'tricky trap' states $\ket{\bm{X}^{trap}}$, the subroutine $\mathcal{K}$ is implemented here, via a controlled operation $\ket{0}_{k}\prescript{}{k}{\bra{0}}\otimes \mathbb{I}_{d+1}+\ket{1}_k\prescript{}{k}{\bra{1}}\otimes E$ with $E=diag(-1,1,1,\cdots,1)$. Therefore, the full output state before $R_x(\eta)$ is
\begin{widetext} 
\begin{eqnarray}
\!\cos{\eta}\ket{0}_k \ket{0}_{up}\ket{0}_{d} \ket{\bm{X}}\!+ \!\sin{\eta}\ket{0}_k\ket{1}_{up}\ket{0}_{d}\mathcal{K}\mathcal{D}\ket{\bm{X}}\!+\!\sin{\eta} \ket{0}_k\ket{1}_{up}\ket{1}_{d} \mathcal{D}^{\perp}\ket{\bm{X}}\!+\!\sin{\eta} \ket{1}_k\ket{1}_{up}\ket{0}_{d}\!(\mathbb{I}\!-\!\mathcal{K}\!)\mathcal{D}\ket{\bm{X}}\!.
\end{eqnarray}
\end{widetext} 

After $R_x(\mp\eta)$ and the post selection which is on the state $\ket{0}_k\ket{0}_{d_1}\ket{0}_{d_2}$ being applied, only the state
\begin{eqnarray}\label{output}
\ket{\bm{X}'}\propto\cos^2{\eta}\ket{\bm{X}}\pm\sin^2{\eta} \mathcal{K} \mathcal{D} \ket{X}
\end{eqnarray}
is remained with a package error $\mathcal{E}=\epsilon_{p}+\epsilon_{pca}+\epsilon_{hs}$, which is the Eq.(\ref{quantum_update}) with a tunable learning rate $\xi= \tan^2{\eta}$.

The complexity of this stage comes from the implementation of $\mathcal{K}$, which requires $\mathcal{O}(\log{(d+1)})$ \emph{Toffoli} gates and $\mathcal{O}(\log{(d+1)})$ extra qubits\cite{supp}.

\emph{Success probability:} 
In addition, the success probability of the each single iterative step with output $\ket{\bm{X}'}$ in Eq.(\ref{output}) can be specified as
\begin{eqnarray} 
P_{succ}&&=\cos^4{\eta}+\sin^4{\eta}|\mathcal{K}\mathcal{D}\ket{\bm{X}}|^2 \nonumber \\
&&\pm \sin^2{\eta}\cos^2{\eta} (\bra{\bm{X}}\mathcal{K}\mathcal{D}\ket{\bm{X}}+\bra{\bm{X}}\mathcal{D}\mathcal{K}\ket{\bm{X}})) 
\end{eqnarray}
Obviously, in the region $\xi=\tan^2{\eta} \leq 1/2$, we have $P_{succ} \geq \cos^4 \eta -2 \sin^2\eta \cos^2 \eta$. That is, the success probability of each iteration can always be bounded finite by choosing of suitable $\xi$. For example, when we take  $\xi=1/3$, $P_{succ}$ will always be larger than $3/16$.

Noticeably, success probability and multi-copies of the states $\ket{\bm{X}}$ used in \emph{subroutine--3} may hinder the popularization of the protocol, as the memory utilization or the final success probability in  slow-converging problem would bring disastrous impact to our devices and results. From this point of view, the number of iterations should be limited, i.e. a  fast convergence problem is acceptable.

\emph{\textbf{Simulation}---}
To test the performance of our protocol, we explore two intuitionistic problems, including both maximum and minimum cases with corresponding cost functions described as 
\begin{eqnarray}
max \quad f_1 &&=\frac{1}{2} (1,x) ^{\otimes 2}A(1,x)^{T\otimes 2} \nonumber \\
min \quad f_2 &&=\frac{1}{2} (1,x_1,x_2) ^{\otimes 2}B (1,x_1,x_2)^{T \otimes 2} 
\end{eqnarray}
where $A$ and $B$ are coefficient matrix, which are specified in Appendix\cite{supp}. $x$=$4,14$ ($f_1$) and $\bm{x}$=$(\pm5,\pm5)$($f_2$)  were set as the initial inputs. The system was driven by a circuit depicted in Fig.\ref{algorithm_circuit} which includes at most $15$ ancillary qubits($k=1$,$up=1$,$d=1$, $e\leq12$). The simulation is gate-based with the assumption of accessible $e^{iDt}$. Thereby, the variable updates itself in compliance with Eq.(\ref{quantum_update}) iteratively.

\begin{figure}[!ht] 
  \centering 
  \includegraphics[angle=0,width=1\columnwidth]{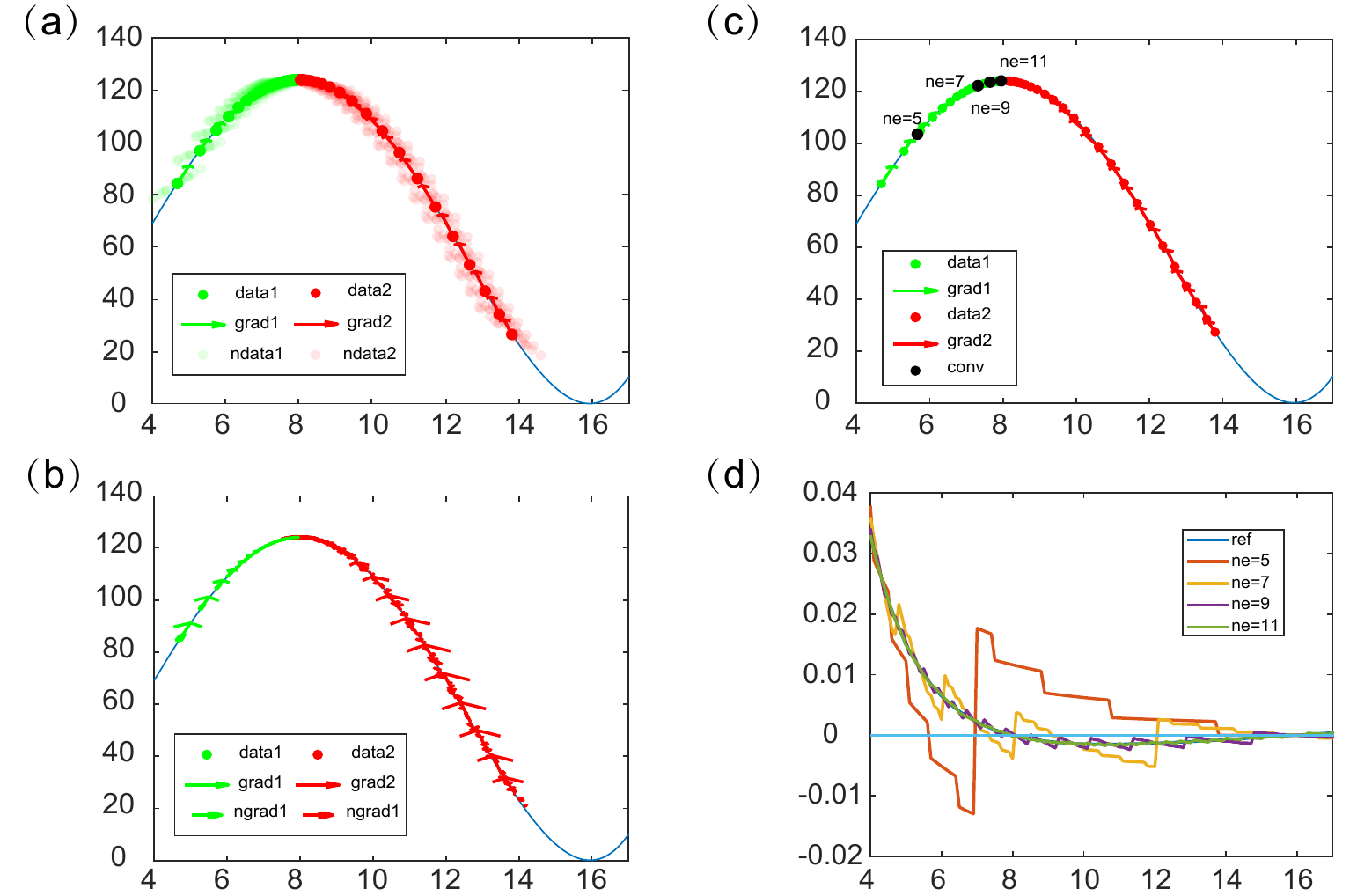} 
  \caption{Numerical simulation of $1$-$d$ optimization. Two projects with different initial trials are labeled in either green or red. Disturbance on the initial guess is shown in (a) with the iterations of the perturbed data(ndata1,ndata2). (b) shows imperfection of the gradient operator, with the slightly disturbed gradient, ngrad1 and ngrad2.
(c) and (d) reveal the results of truncation errors from register $e$. With different sizes of register $e$(ne), the converging point(conv) varies as it encounter the zero-point of the gradient at the first time.}
\label{result1} 
\end{figure}

\begin{figure}[htbp] 
\centering 
\includegraphics[angle=0,width=1\columnwidth]{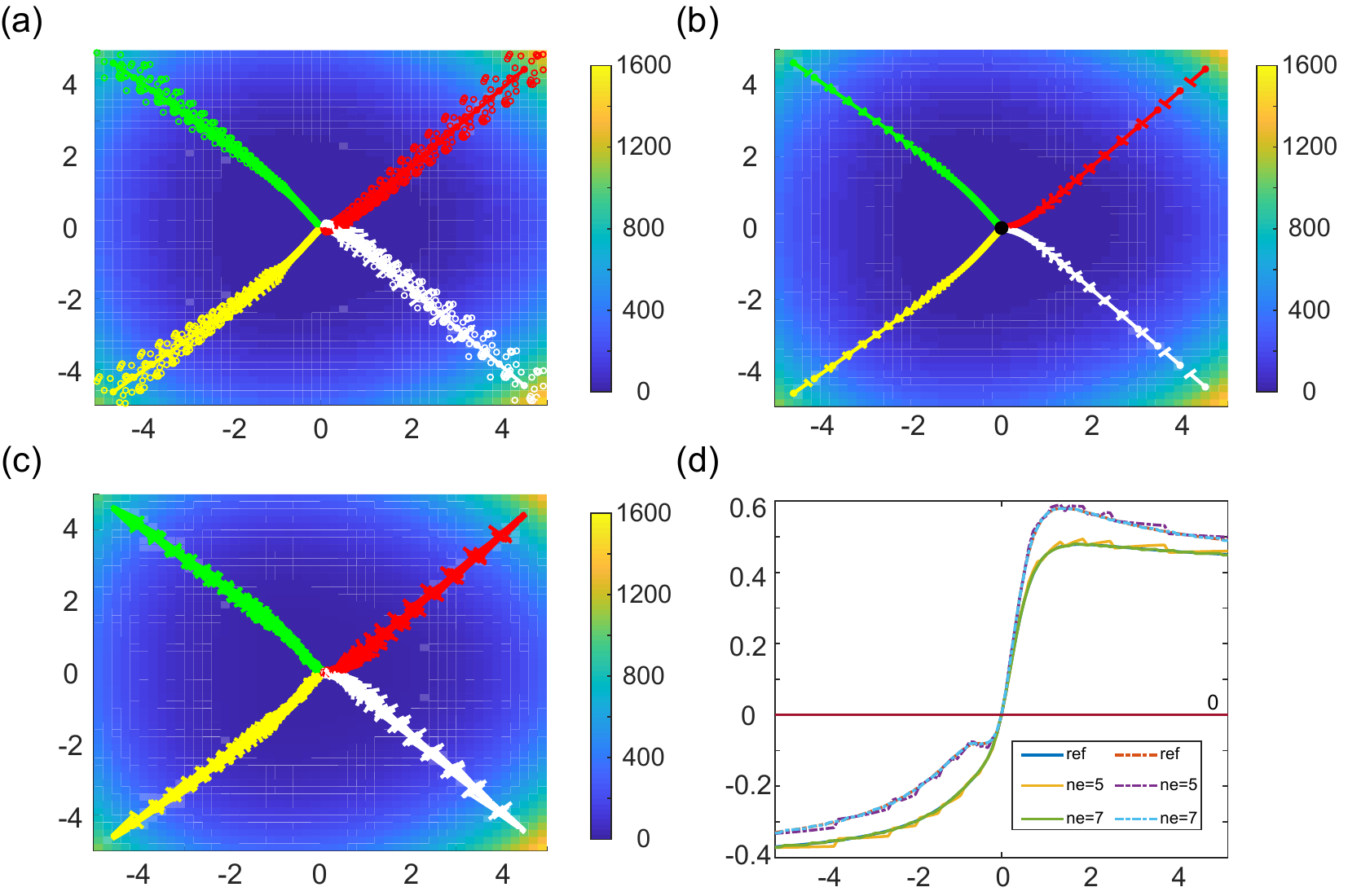} 
\caption{Numerical simulation of $2$-$d$ minimum optimization. Four projects (white, yellow, green, red) are initialized with $\pm(5,5)$, respectively. The insensitivity of the initial perturbation is shown in (a). The imperfectly implemented gradients,which is caused by $\epsilon_{\mathcal{D}}$, are displayed in (b). 
(c) unshrouds that the converging point stays although the different sizes of register $e$(ne), as a rapid climbing of the gradient around the extreme point, which is shown in (d).} 
\label{result2} 
\end{figure}

Results of simulation for optimizing  $f_1$ and $f_2$ are represented in Fig.\ref{result1} and Fig.\ref{result2}, in which $3$ types errors are considered.(1)The initial error $\epsilon_I$, coming from the imperfection of initialization, (2) the operation error $\epsilon_{\mathcal{D}}$, the imperfection when generating $\mathcal{C_{U_D}}$(In real situation, this comes from the \emph{subroutine-3} caused by the defective Hamiltonian simulation or qPCA), and (3) the phase estimation error $\epsilon_{ph}$ for truncation caused by the size of register $e$.

Sub-figures (a) in Fig.\ref{result1} and Fig.\ref{result2} reveal the insensitive consequences of $\epsilon_I$ as the results hold unchanged while initial variables were sampled $20$ times within $5\%$ uniform random distribution around the setting values.

In sub-figures (b),  to simulate the error  $\epsilon_{\mathcal{D}}$, $1\%$ and $5\%$ random perturbations were applied to the $\mathcal{D}$ for $f_1$ and $f_2$, respectively. Obviously, a limited $e_{\mathcal{D}}$ just leads to a slight deviation in both value and direction of the gradient, whose influence would be averaged as unstoppable iterations. However, This influence is problem-dependent. It depends on how close the different extreme points are. if the feasible region is complex, i.e. the different extreme points are too close or $e_{\mathcal{D}}$ is comparably large, the iteration would get to fake extreme points with the stochastic perturbation. Anyway, in all likelihood, the correct optimal result would be finally obtained under a tolerable error $\epsilon_{\mathcal{D}}$.

Sub-figures (c) and (d) show the effects of $\epsilon_{ph}$ by the truncation in register $e$ during the phase estimation. Generally, the size of register $e$ determines the precision of the estimated eigenvalues $\{\lambda_k\}$ and thus the gradient operator. However, in practice, the sensitivity for the size of register $e$ also depends on gradient value near the extreme points. For $f_1$, the truncation error influences the point of convergence heavily as gradient varies slowly around $0$, whose sign of $\pm$ is easily switched. The black points in Fig.\ref{result1}(b) are the convergence points when the size of register $e$($n_e$) is $5$,$7$,$9$,$11$, which indicates that we need choose at least $n_e=12$ to get an acceptable result. On the other hand, for $f_2$, the size of register $e$ affects little to the position of the final result since gradient varies rapidly around the extreme point $(0,0)$. 

\section{\label{sec5:level1}Discussion}

In this paper, we propose an upgrade quantum gradient algorithm for general polynomial optimizations. With the DAE method and the non-unitary $\mathcal{K}$, it demolish the barrier of homogeneous of the cost function\cite{2019-Rebentrost-Qgradient}. As constrained optimization problems usually can be transformed into unconstrained optimization problems, our protocol, without the normalization constrain on argument, extends the framework to more optimization problems. Additionally, compared with the classical counterpart which costs $\mathcal{O}(d)$ operations and $\mathcal{O}(d)$ storage, the time consumption, as well as the memory utilization is reduced to $\mathcal{O}(poly(log(d)))$ in its quantum version. Furthermore, numerical simulation was inspected with the noise and perturbation---initial error $\epsilon_I$, operation error $\epsilon_{\mathcal{D}}$ and phase estimation error $\epsilon_{ph}$.  By the simulation results, the protocol shows its robustness to a tolerant errors, which is important for the real application in NISQ computers. 

However, on a conservative estimation, every iteration the protocol costs multi-copies states due to Hamiltonian simulation and qPCA method.  If the number of iterations required is unlimited, the protocol would be inefficient and unacceptable as the memory consumption grows exponentially. Besides, repeatedly run a probability algorithm would cause success probability decay exponentially. Therefore, only with fast convergence problems or with other combination methods which can get to the feasible region around converging point fast, our algorithm can perform efficiently in both memory and time consumption. 

For modern day architectures, the number of optimized parameters would be at a level of billions. As the world become more and more intelligent, the field of the optimization which is the key to the machine-learning, especially in training the ML model whose objective function is polynomials, is requiring their heavy-lift to process more and more high-dimensional data. Although our protocol cannot give a clean answer to the optimization problem, even to a general polynomials, it is still significant for this field in certain cases. In the case when the feasible region is not comparably huge for the searching point, that means the iteration required is not too much, this quantum gradient method may provide exponential improvements over their classical counterparts.

\textbf{Acknowledgements---}
K.L. acknowledge the National Natural Science Foundation of China under grant Nos. 11905111.  P.G. J.G. S,W. and G.L. acknowledge the National Natural Science Foundation of China under Grants No. 11974205, and No. 11774197. The National Key Research and  Development Program of China (2017YFA0303700); The Key Research and  Development Program of Guangdong province (2018B030325002); Beijing Advanced Innovation Center for Future Chip (ICFC).

\begin{widetext}
\section{Appendix}

\tableofcontents

\section{\label{sec2:level1} Gradient-based Algorithms}
Gradient algorithms, a first-order method, is widely applied in problems such as
\begin{eqnarray}
\max_{\bm{x}} f(\bm{x}) \quad \mbox{or} \quad  \min_{\bm{x}}  f (\bm{x}) , \quad \mbox{for all} \quad \bm{x} \in \mathbb{R}^d,
\end{eqnarray}
where we confine the $d$-dimension variable $\bm{x}$ to the real 
since the complex numbers can be processed with two independent real ones. Thus the problems can be solved iteratively with an iterative equation which is defined as
\begin{eqnarray}
\bm{x}^{t+1}=\bm{x}^t \pm \xi \bm{\nabla} f(\bm{x}^t),
\end{eqnarray}
where $\xi$ is dubbed as the learning rate, the superscript $^t$ is labelled as the $t$-th step of iteration and $+(-)$ corresponds to the maximum(minimum) problems, respectively.

Various algorithms, derived from the prototypical gradient method, is playing an important role in nowadays machine-learning technology. Those gradient-based algorithms include fist-order methods such as Vanilla, Stochastic, Mini-batch, Momentum, RMSprop and Adam methods, and the second-order methods such as Newton\cite{2019-Zhang-GforDP}. However, they are suffering the unmanageable consumption of either time or space for high-dimension parameter optimization since the typical gradient vector $\bm{\nabla}f(\bm{x})$ in step $t$ is usually approximated by a numerical differentiation methods as
\begin{eqnarray}
\frac{\partial _i f(\bm{x})}{\partial x_i} |_{\bm{x}^t}= \lim_{\delta \rightarrow 0}\frac{f( \cdots,  x^t_i +\frac{\delta}{2} ,\cdots)
-f(\cdots, x^t_i -\frac{\delta}{2} ,\cdots)} { \delta}.
\end{eqnarray}
This is known as symmetric difference quotient for discarding of second-order term and involves $2d$ times query of Oracle $O_{f}$ which act as $O_f(\bm{x})=f(\bm{x})$. 
So, when the variable dimension $d$ grows larger and larger, especially being in billions for nowadays machine-learning, extremely high cost of time and space is required and the method is thus out of efficiency.

\section{\label{sec3:level1} Framework of Quantum Gradient Descent Algorithm}
\subsection{\label{Binary_encoding}Binary coding quantum algorithm}
We review the framework of quantum gradient descent method proposed by Jordan, who encodes the variable in the binary from\cite{2005-jordan-Qgradient} 
\begin{eqnarray}
\bm{x}^t \rightarrow \ket{\bm{x}^t}=\ket{x^t_1}\otimes \ket{x^t_2} \otimes \cdots \otimes \ket{x^t_d} .
\end{eqnarray}
To assist the implementation of the gradient method, an Oracle $U_f$ which act as
\begin{eqnarray}\label{eqOf}
U_f: \quad U_f\ket{\bm{x}^t}=e^{i \frac{2\pi}{N_o}\frac{\eta}{2} f(\bm{x}^t)} \ket{\bm{x}^t}
\end{eqnarray}
is defined and could be realized by phase-kicking with ancilla qubits in state $\ket{\psi}=\frac{1}{\sqrt{2N_o/\eta}}\sum_{a=0}^{\frac{2N_o}{\eta} -1}e^{-i \frac{2\pi}{N_o} \frac{\eta}{2} a} \ket{a}$ as 
\begin{eqnarray}
 &&\ket{\bm{x}^t}\otimes \ket{\psi} 
 \xrightarrow{f}  \ket{f(\bm{x}^t)} \otimes \ket{\psi} \nonumber \\
 \xrightarrow{C^{-1}_{not}}&&\ket{f(\bm{x}^t)}\otimes \frac{1}{\sqrt{2 N_o/ \eta}}\sum_{a=0}^{2N_o/\eta -1} e^{i\frac{2\pi}{2N_o/\eta}a}\ket{a-f(\bm{x}^t)}   \nonumber \\
\xrightarrow{f^{-1}}&&e^{i\frac{2\pi}{N_o} \frac{\eta}{2}f({\bm{x}^t})}\ket{\bm{x}^t}\otimes \ket{\psi}
\end{eqnarray}
And also, many other type modular addition can be done by modified control-Not gates, that is, phase kicking is a flexible method which is generally used to exponentiate binary numbers as phase value in quantum computation world. With the hand of those, the full algorithm is given conditionally on the ancillary qubits as 
\begin{eqnarray}
initial:\quad &&\ket{\bm{x}^t} \otimes \ket{0}   \nonumber \\
 \xrightarrow{Hardmard}\quad &&\sum_{\bm{\delta}}\ket{\bm{x}^t} \otimes \ket{\bm{\delta}}   \nonumber \\
 \xrightarrow{C_{not}}\quad &&\sum_{\bm{\delta}}\ket{\bm{x}^t+ \xi \bm{\delta}} \otimes \ket{\bm{\delta}}   \nonumber \\
 \xrightarrow{U_f}\quad &&\sum_{\bm{\delta}}e^{i\frac{2\pi}{N_0} \frac{\eta}{2} f(\bm{x}^t+\xi \bm{\delta})} \ket{\bm{x}^t+ \xi \bm{\delta}} \otimes \ket{\bm{\delta}}   \nonumber \\
 \xrightarrow{cnot^{-2}}\quad &&\sum_{\bm{\delta}}e^{i\frac{2\pi}{N_0} \frac{\eta}{2} f(\bm{x}^t+\xi \bm{\delta})} \ket{\bm{x}^t- \xi \bm{\delta}} \otimes \ket{\bm{\delta}}   \nonumber \\
\xrightarrow{U_f^{-1}}\quad &&\sum_{\bm{\delta}}e^{i\frac{2\pi}{N_0} \frac{\eta}{2} [f(\bm{x}^t+\xi \bm{\delta})-f(\bm{x}^t-\xi \bm{\delta})]} \ket{\bm{x}^t- \xi \bm{\delta}} \otimes \ket{\bm{\delta}}   \nonumber \\
\xrightarrow{cnot}\quad &&\sum_{\bm{\delta}}e^{i\frac{2\pi}{N_0} \eta \xi \bm{\delta} \cdot \bm{\nabla} f(\bm{x}^t)} \ket{\bm{x}^t} \ket{\bm{\delta}} \nonumber \\
=\quad &&\ket{\bm{x}^t} \ket{ \eta \xi \bm{\nabla}f(\bm{x}^t)} \nonumber \\
\xrightarrow{cnot}\quad &&\ket{\bm{x}^t+\eta \xi \bm{\nabla}f(\bm{x}^t)}  \ket{\eta \xi \bm{\nabla}f(\bm{x}^t)}
\end{eqnarray}
where $\eta$ scales the step size while $\xi$ should be small enough for efficiency approximation of the gradient and mitigating the effect from non-linear term.

Bit-consumption is exactly the same as the classical methods. To preserve a $d$-dimension real variable to an accuracy $2^{-\chi}$, $d\chi$ qubits are required. As for the gate complexity, $O(d+log_{2}d)$ quantum gates are required from quantum Fourier transformation, just maintaining a same level as the classical algorithm. However, query complexity is reduced to only 2 for calling oracle $O_f$ while $2d$ queries are needed in classical situation.

\subsection{\label{Amplitude_encoding} Amplitude coding quantum algorithm for even homogeneous polynomials}
Variables are usually encoded as qubit-amplitude in quantum information processing for resources reduction and speed up. Rebentrost et al, absorbed it in their original quantum gradient descent work, which dealing with $2p$-order homogeneous polynomials\cite{2019-Rebentrost-Qgradient}. This optimization problem of homogeneous polynomials can be  expressed in tensor language as
\begin{eqnarray}\label{object_f}
 f(\bm{x}^t)&&=\sum_{m_1+\cdots+m_d=2p} a_{m_1,\cdots,m_d}(x_1^t)^{m_1}\cdots(x_d^t)^{m_d} \nonumber \\
&&=\frac{1}{2} \bm{x}^{T\otimes} A \bm{x}^{\otimes p} \nonumber \\
&&=\frac{1}{2}\sum_{\alpha=1}^K \prod_{n=1}^{p} (\bm{x}^t)^T A_n^{\alpha} \bm{x}^t
\end{eqnarray}
where $a_{m_1,\cdots,m_d}$ is the coefficient of polynomials $f$ and $\bm{x}^t=(x_1^t,x_2^t,\cdots,x_d^t)^T$ is the multi-dimension variable. By involving the coefficients in a matrix $A_{d^p \times d^p}$, $f$ can be seen as matrix product in conjugate of $A$ and multi-folder vector $\bm{x}^{\otimes p}$, as in the second line in Eq.(\ref{object_f}). $A$ can always be chosen be real symmetric and composed of a sum of tensor product of unitary matrices $A=\sum_{\alpha=1}^K (\otimes_{n=1}^p A_n^{\alpha})$, where $A_n^{\alpha}$ acts on the $n$-th $d$-dimension variable space. 

With above depiction, the gradient of $f(\bm{x}^t)$ can be explicitly determined by the current position $\bm{x}^t$
\begin{eqnarray}
\bm{\nabla} f(\bm{x})=\sum_{\alpha=1}^{K}\sum_{n_1=1}^{p} A_{n_1}^{\alpha} \bm{x}^t\prod_{n_2=1,n_2\ne n_1}^p (\bm{x}^t)^T A_{n_2}^{\alpha} \bm{x}^t
\end{eqnarray}
and the corresponding quantum amplitude encoding expressed as
\begin{eqnarray}
\bm{x}^t=(x_1^t,x_2^t,\cdots,x_{d}^t)^T   
\rightarrow \ket{x}^t=\sum_{i=1}^{d}x_i^t\ket{i},
\end{eqnarray}
which implies the normalization constraint $||\bm{x}^t||=1$. That is, the variable is constrained on a $d$-dimension sphere. It benefits the case where the results need not to be expressed explicitly but only used for further analysis, such as expectation value of some certain observables. In this way,  the objective function as expectation value and gradient state can be rewritten as 
\begin{eqnarray}
f(\bm{x}^t)=\frac{1}{2}\sum_{\alpha=1}^K \prod_{n=1}^p \bra{\bm{x}^t} A_n^{\alpha} \ket{\bm{x}^t} 
\end{eqnarray}
\begin{eqnarray}
\ket{\nabla f(\bm{x^t})}&& \propto \sum_{\alpha=1}^{K}\sum_{n_1=1}^{p} \prod_{\substack{n_2=1, \\ n_2\ne n_1}}^p \bra{\bm{x}^t} A_{n_2}^{\alpha} \ket{\bm{x}^t} A_{n_1}^{\alpha} \ket{\bm{x}^t}=\mathcal{D}_{\bm{x}^t} \ket{\bm{x}^t}
\end{eqnarray}
The key to obtain the iterative state is to apply the operator $\mathcal{D}_{\bm{x}^t}$ to the current state. Generally, it is a position-dependent operator which could be unitary or not. With the help of \emph{Hamiltonian simulation}, \emph{qPCA} and \emph{HHL}-like methods at a cost of extra multi-copies of $\ket{\bm{x}^t}$, this operator application can be achieved as shown latter.

However, multi-copies state are required in each iteration, algorithm complexity would scale exponentially in the number of iterations steps performed. On the other hand, amplitude encoding may perform better in resources consumption, as the memory consumption is $O\left((poly(p)\log{d})^{n_0}\right)$, where $n_0$ is the number of iterations. In cases where a reasonable solution can be obtained with a limited iteration, it is acceptable for the memory consumption with $O(poly \log{d})$. As for the gates complexity or query complexity, we leave them in our following work and analyze them as comparison.

In a word, for this method, the resources consumption scales logarithmically with variable dimension $d$ and exponentially with iteration steps. Thus, it performs well with large $d$ and fast-convergence problems.

\section{Subroutines}
After inspecting the three stages in our algorithm in the main text,  four stubborn problems are left to four subroutines. Here, we will show the corresponding details explicitly, including \emph{Subroutine-1}, for initialization of the quantum register;\emph{Subroutine-2}, for applying the \emph{HHL}-like method;\emph{Subroutine-3}, for implementing the variable-dependent $e^{-i\mathcal{D}}$ with quantum principle component analysis method and \emph{Subroutine-4}, for constructing the Non-Unitary operation $\mathcal{K}$. In this section, we have drop the \emph{step}-subscript or -superscript such as $^t$ and $_t$, and the \emph{variable}-subscript $_{\ket{\bm{X}}}$ for convenience. 

\subsection{Subroutine-1: initialization}

\emph{Subroutine-1} serves as an operator, which drives the formatted variable register $v$ to the current variable state $\ket{\bm{X}}=\cos{\gamma}(\ket{0}+\sum_{i=1}^d x_i \ket{i})$. Typically, the concerned optimal problems are usually insensitive to the initial variables, so we can start with some product state, eg., $\ket{0}^{\otimes \log{(d+1)}}$. Thus, the complexity of this subroutine can be ignored in the first iteration step. As for the following iterations, output of the last iteration can play the role of this subroutine. 

Even in some cases where we have to start with some particular state $\ket{\psi}$, the recent-published result shows only  $\mathcal{O}(poly\log d)$  times of quantum manipulations are required with the help of 'bucket brigade' architecture of quantum random access memory(qRAM) per memory call\cite{2008-giovannetti-Qram,2012-Wu-RobustQRAM}.  An alternative way that based on Grover search can also generate arbitrary quantum state with suitably bounded amplitude with fidelity nearly to 1 by consuming $\mathcal{O}(poly\log d)$ qubits\cite{2006-soklakov-SPboundamp}. Hence a conservative estimate of \emph{Subroutine-1} is admitted within $\mathcal{O}(poly\log d)$ time complexity with error assumed as $\epsilon_{I}$ in this process.

\subsection{Subroutine-2: construction of $D$}
There are two quantum phase estimation modules and one double control operation in \emph{Subroutine-2}. Suppose we have accessible controlled-unitary operation $C_{\mathcal{U_D}}=\ket{1}\bra{1}\otimes \sum_{j=0}^{2^{\chi}-1} \ket{j}\bra{j} e^{-i \frac{2\pi}{2^{\chi}}\mathcal{D}j}+\ket{0}\bra{0}\otimes \mathbb{I}$ (whose implementation will be specified in the next \emph{Subroutine-3}), non-unitary operation $\mathcal{D}$ can usually be constructed by a $HHL$-like method. 
After a single qubit rotation on register $up$ which functions as
\begin{eqnarray}
    \ket{0}_{up} \ket{0}_{d} \ket{0}_e \ket{\bm{X}}_v
    \xrightarrow{R^x_{up}(\eta)}&&(\cos{\eta}\ket{0}_{d}+\sin{\eta}\ket{1}_{up})\ket{0}_{d}\ket{0}_e \ket{\bm{X}}_v,
\end{eqnarray}
this method is sketched with the following derivation.

\emph{First}, the binary estimation of the eigenvalues of $\mathcal{D}$ can be resolved as $\lambda _k$(with corresponding eigenstate $\ket{k}$) to the eigenstate register $e$ by quantum phase estimation method. The Hadamard gates $H_e$, control unitary evolution  $C_{\mathcal{U_D}}$ and controlled quantum Fourier transformation on $e$, $C_{\mathcal{U}^{-1}_F}$(functioned at state $\ket{1}_{up}$) are employed here in sequence. We summarize this procedure as 
\begin{eqnarray}
\xrightarrow{H_e}&&\frac{1}{\sqrt{2^{\chi}}}(\cos{\eta}\ket{0}_{up}+\sin{\eta}\ket{1}_{up})\ket{0}_{d}\sum_{j=0}^{2^{\chi}-1}\ket{j}_e \ket{\bm{X}}_v\nonumber \\
\xrightarrow{C_{\mathcal{U_D}}}&&\frac{\cos{\eta}}{\sqrt{2^{\chi}}}\ket{0}_{up}\ket{0}_{d}\sum_{j=0}^{2^{\chi}-1}\ket{j}_e \ket{\bm{X}}_v
+\frac{\sin{\eta}}{\sqrt{2^{\chi}}}\ket{1}_{up}\ket{0}_{d}\sum_{k=1}^{d+1}\sum_{j=0}^{2^{\chi}-1}\ket{j}_{e} e^{-i\frac{2\pi}{2^\chi}\lambda_k j} \beta_k\ket{k}_v\nonumber \\
\xrightarrow{C_{\mathcal{U}^{-1}_F}}&&\frac{\cos{\eta}}{\sqrt{2^{\chi}}}\ket{0}_{up}\ket{0}_{d}\sum_{j=0}^{2^{\chi}-1}\ket{j}_e \ket{\bm{X}}_v +\frac{\sin{\eta}}{\sqrt{2^{\chi}}}\ket{1}_{up}\ket{0}_{d}\sum_{k=1}^{d+1}\ket{\lambda_k}_e \beta_k\ket{k}_v
\end{eqnarray}

\emph{Then}, a multi-control rotation depends on both register $up$ and $e$, with angle $\theta=\arccos{\lambda_k}$ is applied on register $d$.
\begin{eqnarray} {\label{eigenvalue_rotation}}
\xrightarrow{C_{R^x(\theta)}}&& \frac{\cos{\eta}}{\sqrt{2^{\chi}}}\ket{0}_{up}\ket{0}_{d}\sum_{j=0}^{2^{\chi}-1}\ket{j}_e \ket{\bm{X}}_v+\frac{\sin{\eta}}{\sqrt{2^{\chi}}}\ket{1}_{up}\sum_{k=1}^{d+1}(\lambda_k\ket{0}_{d}+\sqrt{1-\lambda_k^2}\ket{1}_{d})\ket{\lambda_k}_e \beta_k\ket{k}_v
\end{eqnarray}

\emph{Finally}, the inverse conditional phase estimation procedure is conducted and the register $e$ is uncoupled from the working system. 
\begin{eqnarray}
\xrightarrow{C_{\mathcal{U}_F}}&& \frac{\cos{\eta}}{\sqrt{2^{\chi}}}\ket{0}_{up}\ket{0}_{d}\sum_{j=0}^{2^{\chi}-1}\ket{j}_e \ket{\bm{X}}_v+\frac{\sin{\eta}}{\sqrt{2^{\chi}}}\ket{1}_{up}\sum_{k=1}^{d+1}(\lambda_k \ket{0}_{d}+\sqrt{1-\lambda^2_k}\ket{1}_{d})\sum_{j=0}^{2^{\chi}-1}\ket{j}_e e^{-i\frac{2\pi}{2^\chi}\lambda_k j} \beta_k \ket{k}_v\nonumber \\
\xrightarrow{H_e \cdot  C^{-1}_{\mathcal{U_D}}}&& \cos{\eta}\ket{0}_{up}\ket{0}_{d} \ket{0}_e  \ket{\bm{X}}_v+\sin{\eta}\ket{1}_{up}\sum_{k=1}^{d+1}(\lambda_k \ket{0}_{d}+\sqrt{1-\lambda^2_k}\ket{1}_{d})\ket{0}_e \beta_k\ket{k}_v
\end{eqnarray}

A following rotation $R^{x}_{d_1}(\eta)$(or its inverse $(R^{x}_{up}(\eta))^{-1}$) on register $up$ with post selection of $\ket{0}_{up}\ket{0}_{d}$ will reduce the result above to be effective $\frac{1}{\mathcal{N}}\left( \ket{\bm{X}}_v-\tan^2{\eta} \mathcal{D} \ket{\bm{X}}_v \right)$ (or $\frac{1}{\mathcal{N}}\left( \ket{\bm{X}}_v+\tan^2{\eta} \mathcal{D} \ket{\bm{X}}_v \right)$ ), that is, the corresponding update of gradient descent(or ascent). To be noticed that for the symmetry of this subroutine, all the operation dependency on register $up$ except $C_{R^x(\theta)}$ in eq.(\ref{eigenvalue_rotation}) can be omitted with the result unchanged, thus it will make a more flexible realization.

When we talk about the complexity, you will find that $C_{R^x(\theta)}$ can be implemented within $\chi$ two-qubit control rotations. So, with the available $C_\mathcal{U_D}$(the complexity of whose implementation will be discussed in the next subroutine), remaining resources consuming of this subroutine comes from standard \emph{phase estimation} shown as,

\emph{\textbf{Memory utilization}} 
\begin{eqnarray}
\chi=n_p+ \ulcorner log(2+\frac{1}{2\delta_{P}})\urcorner
\end{eqnarray}
qubits are required for binary storing of $\{\lambda_k\}$ with accuracy $\epsilon_p=2^{-n_p}$ and lower bound successful probability $1-\delta_{P}$. 

\emph{\textbf{Time consumption}} 
\begin{eqnarray}
 \mathcal{O}(\chi^2)   \mbox{ gates}
\end{eqnarray}

\subsection{Subroutine-3: Quantum principle component analysis}
In this subroutine, we show that the variable dependent operator $\sum_{j=0}^{2^\chi -1} \ket{j}\bra{j} e^{-i \mathcal{D} \frac{2\pi}{2^\chi} j}$ can be efficiently implemented within error $\epsilon_{\mathcal{D}}$ by \emph{Quantum Principle Component Analysis(qPCA)} methods.

First of all, we denote 
\begin{eqnarray}
\mathcal{M}=\sum_{k=1}^p  P_k A P^{\dagger}_k=\sum_{\alpha=1}^{K}\sum_{n_1=1}^{p} A_{n_1}^{\alpha}  \otimes (\bigotimes_{\substack{n_2=1, \\ n_2\ne n_1}}^p  A_{n_2}^{\alpha})
\end{eqnarray}
and $\rho=\ket{\bm{X}} \bra{\bm{X}}$, where $P_k$ be the permutation between the $1$-st and $k$-th subsystem.

By  simple derivation of \emph{qPCA} process, one can find that $\tau$-time evolution of the variable dependent gradient operator $\mathcal{D}=Tr_{p-1}[\rho^p \mathcal{M}]$ (the partial trace operated on the $p-1$ subsystem except the first one) can be approximated as
    \begin{eqnarray}\label{long_time_evolution}
    &&\underbrace{Tr_{p-1}\left[e^{-i \mathcal{M} \frac{\tau}{m}} \rho ^{\otimes p-1} Tr_{p-1}\left[e^{-i \mathcal{M} \frac{\tau}{m}}  \rho ^{\otimes p-1}\cdots
    Tr_{p-1}[  e^{-i \mathcal{M} \frac{\tau}{m}} \rho ^{\otimes p} e^{i \mathcal{M} \frac{\tau}{m}}]\cdots e^{i \mathcal{M}\frac{\tau}{m}} \right] e^{i \mathcal{M}\frac{\tau}{m}} \right]}_{m - \mbox{Trace}} \nonumber \\
    &&=(e^{-i \mathcal{D} \frac{\tau}{m}})^m \rho (e^{i \mathcal{D} \frac{\tau}{m}})^m + \mathcal{O}(m||\mathcal{D}||_{max}^2 \frac{\tau^2}{m^2}) \nonumber \\
    &&= e^{-i \mathcal{D}\tau} \rho e^{i \mathcal{D} \tau}+ \mathcal{O}(\frac{p^2 ||A||_{max}^2\tau^2}{m})
    \end{eqnarray}

In a further step, $e^{-i\mathcal{M} \frac{\tau}{m}}=\prod_{k=1}^p P_k e^{-i A \frac{\tau}{m}} P_k+ \mathcal{O}(\frac{p^2||A||^2\tau^2}{m^2})$(exactly the same error as in eq.(\ref{long_time_evolution}), thus can be merged into each other latter) and $e^{-i A \frac{\tau}{m}}$ can be easily constructed with only access to the coefficient matrix $A$, that is, the well known two oracles
\begin{description}
\item[Oracle 1] $O_1 \ket{j,k}\ket{0}=\ket{j,k}\ket{A_{jk}}$
\item[Oracle 2] $O_2 \ket{j,l}=\ket{j,k_A(j,l)}$
\end{description}
where $k_A(j,l)$ denotes the column index of $l$-th nonzero element in $A$'s $j$-th row.  

By the results of Childs' and Low's work\cite{2015-DBerry-HSimulation,2017-chuang-Hsimulation}, one can efficient implement the sparse Hamiltonian simulation $e^{-i A \frac{\tau}{m}}$ with error $\tilde{\epsilon}$ within 
$\mathcal{O}(\frac{s||A||_{max}\tau}{m} + \frac{\log{1/\tilde{\epsilon}}}{\log{\log{1/\tilde{\epsilon}}}})$ times queries of $O_1$and $O_2$, combined with $\mathcal{O}(p\log{(d+1)}+ \chi' poly\log{\chi'})$ gates, where $2^{-\chi'}$ be the Binary accuracy of $A$'s element as specified in $O_1$ and $s$ be $A$'s sparsity. The details can be find in next section and  \cite{2015-DBerry-HSimulation,2017-chuang-Hsimulation}. So in a word, this shows that variable dependent $\mathcal{C_{U_D}}$ can be implemented with the help of \emph{qPCA} , \emph{Hamiltonian simulation} and multi copies of states $\rho=\ket{\bm{X}}\bra{\bm{X}}$. In the following, we will discuss about the corresponding complexity and operation errors. 

Noticed that the control evolution $\mathcal{C_{U_D}}=\sum_{j=0}^{2^\chi-1} \ket{j}\bra{j} e^{-i \mathcal{D} \frac{2\pi}{2^\chi} j}$ can be constructed by Eq.(\ref{long_time_evolution}) when $\mathcal{C_{U_A}}=\sum_{j=0}^{2^\chi-1} \ket{j}\bra{j} e^{-i A \frac{2\pi}{m 2^\chi} j}$ takes the place of $e^{-i A \frac{\tau}{m}}$ as mentioned before. So once $\mathcal{C_{U_A}}$ be available, we take $m=\frac{4\pi^2p^2 ||A||_{max}}{\epsilon_{pca}}$ to bound the \emph{qPCA} induce error in $\mathcal{C_{U_D}}$ as $\epsilon_{pca}$. Meanwhile, 
\begin{eqnarray}
\mathcal{O}(mp)=\mathcal{O}(\frac{4\pi^2 p^3||A||^2_{max}}{\epsilon_{pca}})
\end{eqnarray} 
copies of state $\rho=\ket{\bm{X}}\bra{\bm{X}}$ and
\begin{eqnarray}
\mathcal{O}(\frac{4 \pi^2 p^3 ||A||^2_{max} \log{(1+d)}}{\epsilon_{pca}})
\end{eqnarray}
two qubits swap gates(for implementation of $P_k$), required in this \emph{qPCA} process.

Besides, since the $\mathcal{C_{U_A}}$ can be constructed as
\begin{eqnarray}\label{control_decomp}
\sum_{j=0}^{2^\chi-1} \ket{j}\bra{j} e^{-i A \frac{2\pi}{m 2^\chi} j}= \sum_{\{j_k=0,1\}} \ket{j_1}\bra{j_1} e^{-i A\frac{2\pi}{m2} j_1} \cdots \ket{j_k}\bra{j_k} e^{-i A \frac{2\pi}{m 2^k} j_k} \cdots \ket{j_\chi}\bra{j_\chi} e^{-i A \frac{2\pi}{m 2^\chi} j_\chi}
\end{eqnarray}   
and the Hamiltonian simulation complexity scales linearly with evolution time, usually in the case of limited $\xi$, the complexity in constructing $\mathcal{C_{U_A}}$ contributes to the 1st term in Eq.\ref{control_decomp}. We can easily find that $mp$ times queries of $\mathcal{C_{U_A}}$ share the same implementation complexity and Hamiltonian simulation induced error $\epsilon_{hs}$ with $e^{-i A 2p\pi}$, specifically, that is
\begin{eqnarray}
\mathcal{O}(2p\pi s||A||_{max}+\frac{\log{1/\epsilon_{hs}}}{\log{\log{1/\epsilon_{hs}}}})
\end{eqnarray}
times queries of $O_1$ and $O_2$.

In the last, the full error(both from \emph{qPCA} and \emph{Hamiltonian simulation}) of control evolution $\mathcal{C_{U_D}}$ can be bounded by these two independent error as
\begin{eqnarray}
\epsilon_{\mathcal{D}}=\epsilon_{pca} + \epsilon_{hs}
\end{eqnarray}

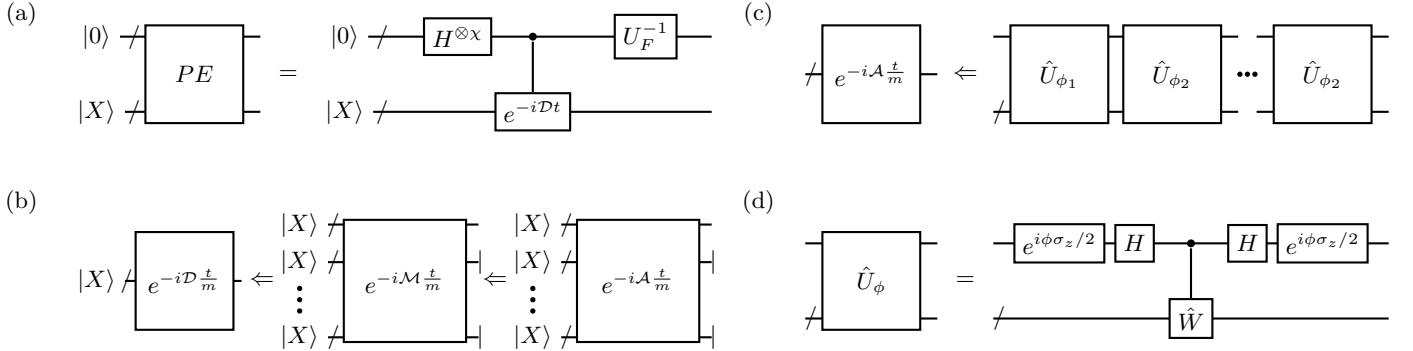
\begin{figure}[!h]
    \centerline{
    \begin{tikzpicture}[thick]
      \tikzstyle{operator} = [draw,fill=white,minimum size=1.5em] 
      \tikzstyle{operator2} = [draw,fill=white,minimum size=4em] 
      \tikzstyle{operator3} = [draw,fill=white,minimum size=5em] 
      \tikzstyle{phase} = [fill,shape=circle,minimum size=3pt,inner sep=0pt]
      \tikzstyle{surround} = [fill=blue!10,thick,draw=black,rounded corners=2mm]
      \tikzstyle{ellipsis} = [fill,shape=circle,minimum size=2pt,inner sep=0pt]
      \node at (0.2,0.3) (note) {(a)};
      \node at (1.2,0) (q1) {$\ket{0}$};
      \node at (1.2,-1) (q2) {$\ket{X}$};
      \node at (1.7,0) (hy11) {$/$};
      \node at (1.7,-1) (hy22) {$/$};
      \node (end1) at (3.5,0) {} edge [-] (q1);
      \node (end2) at (3.5,-1) {} edge [-] (q2);
      \node[operator2] (op21) at (2.5,-0.5) {$PE$} ;
      \node at (3.75,-0.5) (eq1) {$=$};
      \node at (4.5,0) (q11) {$\ket{0}$};
      \node at (4.5,-1) (q22) {$\ket{X}$};
      \node at (5,0) (hy1) {$/$};
      \node at (5,-1) (hy2) {$/$};
      \node (end1) at (9.5,0) {} edge [-] (q11);
      \node (end2) at (9.5,-1) {} edge [-] (q22);
      \node[operator] (op2) at (6,0) {$H^{\otimes{\chi}}$};
      \node[phase] (phase11) at (7,0) {};
      \node[operator] (op5) at (7,-1) {$e^{-i\mathcal{D}t}$};
      \node[operator] (op3) at (8.5,0) {$U_F^{-1}$};
      \draw[-] (op5) -- (phase11);
      \node at (0.2,-2.2) (note) {(b)};
      \node at (1.2,-3.25) (q3) {$\ket{X}$};
      \node (end3) at (3.25,-3.25) {} edge [-] (q3);
      \node at (1.6,-3.25) (hy111) {$/$};
      \node[operator2] (op22) at (2.37,-3.25) {$e^{-i\mathcal{D}\frac{t}{m}}$} ;
      \node at (3.4,-3.25) (eq1) {$\Leftarrow$};
      \node at (3.9,-2.5) (q33) {$\ket{X}$};
      \node at (3.9,-3) (q3232) {$\ket{X}$};
      \node at (3.9,-4) (q44) {$\ket{X}$};
      \node (end3) at (6.4,-2.5) {} edge [-] (q33);
      \node (end32) at (6.4,-3) {} edge [-] (q3232);
      \node (end4) at (6.4,-4) {} edge [-] (q44); 
      \node at (4.35,-2.5) (hy33) {$/$};
      \node at (4.35,-3) (hy3232) {$/$};
      \node at (4.35,-4) (hy44) {$/$};
      \node at (6.3,-3) (hy3232) {$|$};
      \node at (6.3,-4) (hy44) {$|$};
      \node[operator3] (op21) at (5.3,-3.25) {$e^{-i\mathcal{M}\frac{t}{m}}$} ;
      \node[ellipsis] (ee1) at (3.9,-3.35) {};
      \node[ellipsis] (ee2) at (3.9,-3.5) {};
      \node[ellipsis] (ee3) at (3.9,-3.65) {};  
      \node at (6.5,-3.25) (eq1) {$\Leftarrow$};
      \node at (7,-2.5) (q33) {$\ket{X}$};
      \node at (7,-3) (q3232) {$\ket{X}$};
      \node at (7,-4) (q44) {$\ket{X}$};
      \node (end3) at (9.5,-2.5) {} edge [-] (q33);
      \node (end32) at (9.5,-3) {} edge [-] (q3232);
      \node (end4) at (9.5,-4) {} edge [-] (q44); 
      \node at (7.45,-2.5) (hy33) {$/$};
      \node at (7.45,-3) (hy3232) {$/$};
      \node at (7.45,-4) (hy44) {$/$};
      \node at (9.4,-3) (hy3232) {$|$};
      \node at (9.4,-4) (hy44) {$|$};
      \node[operator3] (op21) at (8.4,-3.25) {$e^{-i\mathcal{A}\frac{t}{m}}$} ;
      \node[ellipsis] (ee1) at (7,-3.35) {};
      \node[ellipsis] (ee2) at (7,-3.5) {};
      \node[ellipsis] (ee3) at (7,-3.65) {}; 
       \node at (10,0.3) (note) {(c)};
      \node at (10.5,-0.5) (q4) {};
      \node at (10.7,-0.5) (hyC1) {$/$};
      \node (end4) at (12.5,-0.5) {} edge [-] (q4);   
      \node[operator2] (op22) at (11.5,-0.5) {$e^{-i\mathcal{A}\frac{t}{m}}$} ;
      \node at (12.75,-0.5) (eq1) {$\Leftarrow$};
      \node at (13,0) (q33) {};
      \node at (13,-1) (q44) {};
      \node at (13.2,-1) (hyC2) {$/$};
      \node (end3) at (16.5,0) {} edge [-] (q33);
      \node (end4) at (16.5,-1) {} edge [-] (q44);   
      \node at (16.5,0) (q333) {};
      \node at (16.5,-1) (q444) {};
      \node (end3) at (18.5,0) {} edge [-] (q333);
      \node (end4) at (18.5,-1) {} edge [-] (q444);   
      \node[operator2] (op21) at (14,-0.5) {$\hat{U}_{\phi_1}$} ;
      \node[operator2] (op21) at (15.5,-0.5) {$\hat{U}_{\phi_2}$} ;
      \node[operator2] (op21) at (17.5,-0.5) {$\hat{U}_{\phi_2}$} ;
      \node[ellipsis] (ee1) at (16.4,-0.5) {};
      \node[ellipsis] (ee2) at (16.5,-0.5) {};
      \node[ellipsis] (ee3) at (16.6,-0.5) {};    
       \node at (10,-2.2) (note) {(d)};
       \node at (10.5,-2.75) (q1) {};
       \node at (10.5,-3.75) (q2) {};   
       \node at (10.7,-3.75) (hyD1) {$/$};
       \node (end1) at (12.5,-2.75) {} edge [-] (q1);
       \node (end2) at (12.5,-3.75) {} edge [-] (q2);
       \node[operator2] (op21) at (11.5,-3.25) {$\hat{U}_{\phi}$} ;
       \node at (12.75,-3.25) (eq1) {$=$};
       \node at (13,-2.75) (q11) {};
       \node at (13,-3.75) (q22) {};
       \node at (13.2,-3.75) (hyD2) {$/$};
       \node (end1) at (18.5,-2.75) {} edge [-] (q11);
       \node (end2) at (18.5,-3.75) {} edge [-] (q22);
       \node[operator] (op1) at (14,-2.75) {$e^{i\phi\sigma_z/2 }$};
       \node[operator] (op2) at (15,-2.75) {$H$};
       \node[phase] (phase11) at (15.75,-2.75) {};
       \node[operator] (op5) at (15.75,-3.75) {$\hat{W}$};
       \node[operator] (op3) at (16.5,-2.75) {$H$};
       \node[operator] (op4) at (17.5,-2.75) {$e^{i\phi\sigma_z/2 }$};
       \draw[-] (op5) -- (phase11);
    \end{tikzpicture} } 
    \caption{The structure to implement the $subroutine$-$2$.(a) is the standard quantum phase estimation modular, where $e^{-i \mathcal{D} \tau}$ can be simulated in (b), which simulate a piece of time $\tau/m$ with the $subroutine$-$3$, via the qCPA and Hamiltonian simulation of $\mathcal{A}$.(c) and (d) are the Hamiltonian simulation with quantum signal processing method by adding one ancillary qubit.}   
    \label{oracle gradient} 
  \end{figure}

\subsection{Subroutine-4: construction of $\mathcal{K}$}
Recall that since there is a projection constraint $||\bm{x}||=1$ in previous work\cite{2019-Rebentrost-Qgradient}, all variable lied in a high-dimension sphere and the iteration stopped at state $\ket{\bm{x}}\propto \mathcal{D}\ket{\bm{x}}$ since the projected gradient vanished.

However, as the projection constrained moved out in our work by the dressed amplitude encoding $\bm{x}\rightarrow \ket{\bm{X}}=\cos{\gamma}(\ket{0}+\sum_{i=1}^d \ket{i})$,  quantum state $\ket{\bm{X}^{op}}$ which satisfy $\ket{\bm{X}^{op}} \propto \ket{\bm{X}^{op}} - \mathcal{K} \mathcal{D} \ket{\bm{X}^{op}}$ be the new optimized result, instead of state $\ket{\bm{X}^{trap}}$ which satisfy $\mathcal{D}\ket{\bm{X}^{trap}} \propto \ket{\bm{X}^{trap}}$, where $\mathcal{K}=diag(0,1,1,\cdots,1)$. 

Non-unitary $\mathcal{K}$ can be well implemented with the help of one single ancilla qubit as
\begin{flalign}\label{post_selection}
\mathcal{K}\ket{\psi} \propto \ket{0}\bra{0}_a H_a C_E H_a \ket{0}_a\ket{\psi}
\end{flalign}
where $H_a$ be Hadamard and $C_E =\ket{0}\bra{0}_a \otimes \mathbb{I}+ \ket{1}\bra{1}_a \otimes diag(-1,1,1,\cdots,1)$(or in another view, $C_E=C^0_{Z_a}$, a control-Z gate on ancilla functioned when the principle system in state $\ket{0}$). The success probability of post selection on ancilla qubit state $\ket{0}$ equal to the weight of non-$\ket{0}$ conponent in $\ket{\psi}$, which match the applied condition of this optimal method, that is, the object function being fast converged (that is, with finite gradient).

Notice that $C_E=C^0_{Z_a}=X^{\otimes \log{(d+1)}} C^1_{Z_a} X^{\otimes \log{(d+1)}}$, where $C^1_{Z_a}$ can be easily implemented with $\mathcal{O}(\log{(d+1)})$ \emph{Toffoli} gates and $\mathcal{O}(\log{(d+1)})$ extra qubits which set in $\ket{0}$(or in another way, this can be done with $\mathcal{O}(\log^2{(d+1)})$ \emph{Toffoli} gates without ancillary).

\section{Hamiltonian simulation by quantum signal processing}

In this section, the method of quantum signal processing(QSP) to complete the evolution $e^{-iHt}$ of $d$-dimension Hamiltonian $H$( where $A$ play the role of $H$ in\emph{subroutine-3}) is introduced. This method is based on the sparse matrix assumption, which involves two oracles.
\begin{description}
    \item[Oracle 1] $O_1 \ket{j,k}\ket{0}=\ket{j,k}\ket{H_{jk}}$
    \item[Oracle 2] $O_2 \ket{j,l}=\ket{j,k_H(j,l)}$
\end{description}
where polynomial coefficients oracle $O_1$ operates on $\chi'+2\log{(d)}$ qubits with $i,j= 1, \cdots,  d$, to hold an accuracy of $H$'s elements to $2^{-\chi'}$. The function $k_H(j,l)$ output the column index of $l$-th nonzero element in $j$-th row of $H$. Sparse input oracle $O_2$ operates on $2\log{d}$ qubits and $l=1, \cdots, s$, where $s$ just be the sparsity of $H$.

The result shows one can efficient implement the sparse-$s$ Hamiltonian simulation $e^{-iHt}$ with error $\epsilon_{hs}$ within
\begin{description}
    \item[Query complexity] $\mathcal{O}(s||H||_{max}t + \frac{\log{1/\epsilon_{hs}}}{\log{\log{1/\epsilon_{hs}}}})$
    \item[Gate complexity] $\mathcal{O}(\log{d}+ \chi' poly\log{\chi'})$
\end{description}

In the following, we will sketch the QSP for Hamiltonian simulation. Fig.\ref{oracle gradient} (c) and (d) give the basic idea. Both circuits consist of two input, one ancillary qubit  and a workspace.  A series of $\hat{U}_{\phi}$ build up our simulation circuit, where $\hat{U}_{\phi}$(shown in Fig.\ref{oracle gradient}) consists of the Hadamard gates, $z$-rotations and controlled-$\hat{W}$ operation. when the system input is on the eigenvectors $\ket{\lambda}$, i.e. $\hat{W}\ket{\lambda}=e^{\theta_{\lambda}}\ket{\lambda}$, it can be reduced to the single-qubit rotation $R_{\phi}(\theta_{\lambda})$ on the ancilla, where $R_{\phi}(\theta_{\lambda})=e^{-i(\theta_{\lambda}/2)(\sigma_x cos(\phi)+\sigma_y sin(\phi))}$.

A general $e^{ih(\theta_{\lambda})}$ can be approximated with $R_{\phi_1}(\theta_{\lambda})R_{\phi_2}(\theta_{\lambda})...R_{\phi_n}(\theta_{\lambda})$ via optimizing the parameters $\phi_1,\phi_2,...,\phi_n$ in the sequence of the $\hat{U}_{A}$ (shown in  the Fig.\ref{oracle gradient}(c)), which builds up a transformation on the functional workspace
\begin{eqnarray}
 W&=&\sum_{\lambda}e^{i\theta_{\lambda}}\ket{\lambda}\bra{\lambda}\nonumber \\
 \rightarrow V&=&\sum_{\lambda}e^{ih(\theta_{\lambda})}\ket{\lambda}\bra{\lambda}
\end{eqnarray}
When $h(\theta_{\lambda})=-\tau sin(\theta_{\lambda})$, the simulation circuit can be used in the Hamiltonian simulation via quantum walk. 

We will given a brief illustration on Hamiltonian simulation via QSP. Given a $s$-sparse Hamiltonian $H$ acting on $d$ dimension Hilbert space. First of all, an ancillary qubit with the state $\ket{0}$ is appended, expanding the space from $d$ to $2d$, then the entire Hilbert space is duplicated(thus $\mathbb{C}^{2d} \otimes \mathbb{C}^{2d}$). The whole process can be done with the isometry $T$
\begin{eqnarray}
T=\sum_{j=0}^{d-1} \sum_{b \in \{0,1\}}  (\ket{j} \bra{j} \otimes \ket{b} \bra{b}) \otimes \ket{\phi_{j,b}}
\end{eqnarray}
where $\ket{\phi_{j,0}}$ and $\ket{\phi_{j,1}}$ are defined as 
\begin{eqnarray}
&&\ket{\phi_{j,0}}=\frac{1}{\sqrt{s}}\sum_{l \in S_j} \left( \ket{l}  \sqrt{\frac{H_{i,l}^{\*}}{X}}
\ket{0}+\sqrt{1-\frac{|H_{i,l}^*|}{X}}\ket{1} \right) \nonumber \\
&&\ket{\phi_{j,1}}=\ket{0} \ket{1} 
\end{eqnarray}
with $X\geq ||H||_{max}$ and $S_j$ be the set of indices of nonzero elements in column $j$ of $H$. This is a controlled state preparation step, performing on the input $\ket{j}\ket{b}$, to creat $\ket{\phi_{j,b}}$($b=0,1$). In the step,  one query to the oracle $O_1$ and $O_2$ and additional $(\chi'ploylog(\chi'))$ primitive gates are required.
After this stage, the unitary operator of the quantum walk are applied with
\begin{eqnarray}
U=i S (2TT^{\dagger}- \mathbb{I})
\end{eqnarray}
$S$ is swap operation, acting on the former and latter two register as $S \ket{j_1} \ket{b_1} \ket{j_2} \ket{b_2}=\ket{j_2} \ket{b_2} \ket{j_1} \ket{b_1}$ where $j_1,j_2 \in [d]$ and $b_1,b_2 \in \{0,1\}$.  As $U$ corresponds to reflection about $TT^{\dagger}$ followed by $S$, swapping ($2+2\log{d}$)-qubit registers, its query and gate complexities are identical to $T$ up to constant factors. 

Suppose that $\ket{\lambda}$ be $H$'s eigenstates as $H\ket{\lambda}=\lambda\ket{\lambda}$, The quantum walk operator $U$ and its eigenvalues $\mu_{\pm}$ thus satisfy the following rules
\begin{eqnarray}
&&U\ket{\mu_{\pm}}=\mu_{\pm}\ket{\mu_{\pm}}\nonumber \\
&&\ket{\mu_{\pm}}=(1+i\mu_{\pm}S)T\ket{\lambda} , \mu_{\pm}=\pm e^{\pm i \arcsin{(\lambda /X s)}}
\end{eqnarray}
Since 
\begin{eqnarray}
  || T \frac{1-iS}{\sqrt{2}} (iU)^{\tau}\frac{1+iS}{\sqrt{2}}T^{\dagger} -e^{i\tau sin^{-1}(H/||abs(H)||)}||<(||H||/||abs(H)||)^2
\end{eqnarray} 
If $\lambda /X s$ is small enough, applying $\tau=s||H||_{max}t$ steps of the discrete-time quantum walk $U$, we could simulate $t$ time Hamiltonian simulation since\cite{2010-Achilds-HSimulation}.
Therefore, finally, the inverse state preparation $T^{\dagger}$ is performed. For a successful simulation, the output should lie in the original space, and the ancillary qubit should be returned to the state $\ket{0}$. 

The hint is, the nonlinear of the phase factor $arcsin(\lambda/X s)$ making the simulation deviate from the desired value. The transformation via quantum signal processing from $W$ to $V$ is applied here, when $h(\theta_{\lambda})=-\tau sin(\theta_{\lambda})$. It modifies the simulation circuit to a controlled one(the simulation circuit is just the $W$ in above protocol).

\section{Success probability}
By simple derivation, the full output state in circuits in main text before $R_x(\eta)$ acted can be shown as
\begin{eqnarray}
\cos{\eta}\ket{0}_k \ket{0}_{up}\ket{0}_{d} \ket{\bm{X}}+ \sin{\eta}\ket{0}_k\ket{1}_{up}\ket{0}_{d}\mathcal{K}\mathcal{D}\ket{\bm{X}} \nonumber \\
+\sin{\eta} \ket{0}_k\ket{1}_{up}\ket{1}_{d} \mathcal{D}^{\perp}\ket{\bm{X}} +\sin{\eta} \ket{1}_k\ket{1}_{up}\ket{0}_{d}(\mathbb{I}-\mathcal{K})\mathcal{D}\ket{\bm{X}}
\end{eqnarray}
where $\mathcal{D}^{\perp}\ket{\bm{X}}$ denotes the rubbish state $\sum_k \sqrt{1-\lambda^2_k} \beta_k \ket{k}$.  
Obviously, when $R_x(\mp \eta)$ applied and post selected on state $\ket{0}_k\ket{0}_{d_1}\ket{0}_{d_2}$, only the
\begin{eqnarray}
\cos^2{\eta}\ket{\bm{X}} \pm \sin^2{\eta} \mathcal{K} \mathcal{D} \ket{X}
\end{eqnarray}
preserved, with probability
\begin{eqnarray} \label{succ_probability}
P_{succ}&&=\cos^4{\eta}+\sin^4{\eta}|\mathcal{K} \mathcal{D}\ket{\bm{X}}|^2 \pm \sin^2{\eta} \cos^2{\eta} (\bra{\bm{X}}\mathcal{K} \mathcal{D}\ket{\bm{X}}+\bra{\bm{X}}\mathcal{D} \mathcal{K}\ket{\bm{X}}) 
\end{eqnarray}

Typically, when the tunable learning rate $\xi=\tan^2{\eta}$ chosen to be less than $1/2$, we have
\begin{eqnarray}
P_{succ} \geq \cos^4{\eta} -2\sin^2{\eta} \cos^2{\eta}.
\end{eqnarray}
for both descent and ascent cases. That is, the success probability of each iteration can always be bounded as finite by choosing of suitable $\xi$($\eta$). For example, when we take $\xi=1/3$, $P_{succ}$ will always be lager than $3/16$.

\section{Algorithm Simulation}
Based on the theoretical protocol, the simulation program consists of an iteration of three stages as depicted in the article. The program assumed that the Hamiltonian simulation, i.e. $e^{iDt}$ can be simulated directly in the machine. Therefore, two cases are simulated in the frame of our protocol.

The polynomial optimizations include both maximum and minimum problems, which state as 
\begin{eqnarray}
  max \quad f_1 &&=\frac{1}{2} (1,x) ^{\otimes 2}
  \begin{pmatrix}
  7/2 & 0 \\
  0 & -9/2
  \end{pmatrix}^{\otimes 2}
   (1,x)^{T\otimes 2} \nonumber \\
  min \quad f_2 &&=\frac{1}{2} (1,x_1,x_2) ^{\otimes 2}\left[
  \begin{pmatrix}
  1 & 0 & 0\\
  0 & 1 & 0\\
  0 & 0 & 1
  \end{pmatrix}^{\otimes 2} + 
  \begin{pmatrix}
  0 & 0 & 1\\
  0 & 0 & 0\\
  1 & 0 & 0
  \end{pmatrix}\otimes 
  \begin{pmatrix}
  0 & 0 & 0\\
  0 & 0 & 1\\
  0 & 1 & 0
  \end{pmatrix}\right]
  (1,x_1,x_2)^{T \otimes 2} 
\end{eqnarray}

According to the protocol, an arbitrary initial input $(1,\bm{x'})$ can be represented with DAE by a quantum state $\ket{\bm{X}}=\cos{\gamma}(\ket{0}+\sum_i x_i \ket{i})$. All projects are based on the circuit model, which consists of a set of elementary quantum gates, i.e. arbitrary single-qubit rotations and two-qubit controlled-unitary operators. 

Initialization was implemented by the rotations on a qubit or a qutrit, as measurement was realized by simulating tomography on the variable register $e$, after projecting entire system into the target subspace. As for the intermediate procedures, the gates are implemented as the algorithm circuit with the assumption to get $e^{iDt}$. 

Remarkably, the standard phase estimation module approximates the eigenvalue of the target unitary with a range of $(0,1)$. However, this module should be modified since when constructing the $D$, the gradient operator produces both the positive and negative component. The periodic property is utilized in our implementation and the eigenvalues in the gradient operator are normalized in a range of $(-\frac{1}{2},\frac{1}{2})$. Therefore, the output of this module is divided into two parts, results in $(0,\frac{1}{2})$ is directly readout while others in $(\frac{1}{2},1)$ should be resolved into the rangle of $(-\frac{1}{2},0)$ by the periodic condition. 

In the program, the simulation includes $15$ ancillary qubit in a conservative way, with $1$ for $k$, $1$ for $d1$, $1$ for $d2$ and $12$ for $e$, whose resolution has been discussed in article. The rest qubits are used to encode the variables. During the simulation, $\bm{x}=(1,7 \pm 3)$ are considered as the two random initial guess for $f_1$, and $\bm{x}=(1, \pm 5, \pm 5)$ are chosen as four different initial variables for $f_2$. And $0.05$($0.1$) is chosen as the learning rate for $f1$ ($f_2$).

$3$ types errors are investigated, including the initial error $\epsilon_I$, the operation error $\epsilon_{\mathcal{D}}$ and the phase estimation error $\epsilon_{ph}$. $\epsilon_I$ comes from the imperfection of initialization which cannot arrive at the target input. we simulate this situation repeatably for 20 times, by introducing a perturbation with uniform random distribution whose amplitude is $5\%$ to the input state. $\epsilon_{\mathcal{D}}$ is from the uncertainty of the $e^{iDt}$, as $e^{iDt}$ cannot be perfectly generated. We introduced the noise into the operator, with $1\%$ and $2\%$ to the amplitude of $D$ for two cases, and simulated them for 15 times. $\epsilon_{ph}$ is the truncation error which originated from the size of eigenvalue register $e$. To simulate this situation, we choose different sizes of the $e$ for $5$,$7$,$9$,$11$. On the other side, we investigate the effects on gradient value of different sizes of $e$.

\end{widetext}
\bibliographystyle{unsrt}

\end{document}